\documentclass[aps,prl,twocolumn,superscriptaddress,floatfix,showpacs,citeautoscript,longbibliography,hyperlinks]{revtex4-2}
\usepackage{amsmath}
\usepackage{epstopdf}
\usepackage{amsmath}
\usepackage{amssymb}
\usepackage{amsfonts}
\usepackage[colorlinks=true,citecolor=blue]{hyperref}
\usepackage{graphicx,graphics}

\usepackage{float}
\usepackage[caption = false]{subfig}

\newcommand{\revision}[1]{{{#1}}}
\newcommand{\revisiontwo}[1]{{{#1}}}

\begin{document}
\title{Dynamic Cooper Pair Splitter}
\author{Fredrik Brange}
\affiliation{Department of Applied Physics, Aalto University, 00076 Aalto, Finland}
\author{Kacper Prech}
\affiliation{Department of Applied Physics, Aalto University, 00076 Aalto, Finland}
\affiliation{School of Physics and Astronomy, University of Glasgow, Glasgow, G12 8QQ, United Kingdom}
\author{Christian Flindt}
\affiliation{Department of Applied Physics, Aalto University, 00076 Aalto, Finland}

\begin{abstract}
Cooper pair splitters are promising candidates for generating spin-entangled electrons. However, the splitting of Cooper pairs is a random and noisy process, which hinders further synchronized operations on the entangled electrons. To circumvent this problem, we here propose and analyze a dynamic Cooper pair splitter that produces a noiseless and regular flow of spin-entangled electrons. The Cooper pair splitter is based on a superconductor coupled to quantum dots, whose energy levels are tuned in and out of resonance to control the splitting process. We identify the optimal operating conditions for which exactly one Cooper pair is split per period of the external drive and the flow of entangled electrons becomes noiseless. To characterize the regularity of the Cooper pair splitter in the time domain, we analyze the $g^{(2)}$-function of the output currents and the distribution of waiting times between split Cooper pairs. Our proposal is feasible using current technology, and it paves the way for dynamic quantum information processing with spin-entangled electrons.
\end{abstract}

\maketitle

\emph{Introduction.}--- Superconductors are natural sources of entangled particles \cite{Tinkham}. By splitting the Cooper pairs in a supercondutor into different normal-state leads, spin-entanglement between spatially separated electrons can be achieved \cite{Lesovik2001,Recher2001}. Cooper pair splitters have been realized in several types of solid-state architectures \cite{PhysRevLett.93.197003,PhysRevLett.95.027002,Hofstetter:Cooper,PhysRevLett.104.026801,Wei2010,PhysRevLett.107.136801,PhysRevLett.109.157002,Herrmann:Spectroscopy,Das2012,PhysRevB.90.235412,PhysRevLett.114.096602,PhysRevLett.115.227003,Borzenets:High,Bruhat2018,tan2020,PhysRevLett.126.147701,Ranni2021}, for instance using quantum dots \cite{Hofstetter:Cooper,PhysRevLett.107.136801,Herrmann:Spectroscopy}, carbon nanotubes~\cite{PhysRevLett.104.026801}, or graphene \cite{PhysRevLett.115.227003,Borzenets:High,tan2020,PhysRevLett.126.147701}. Experimentally, the splitting process has been observed by measuring the non-local conductance or the noise \cite{Wei2010,Das2012} and recently using single-electron detectors \cite{Ranni2021}. However, with static voltages, the generation of spin-entangled electrons is a random and noisy process, which offers little control over the regularity and the timing of the Cooper pair splitting.

In parallel with these developments, single-electron emitters have emerged as accurate sources of noiseless currents \cite{Feve1169,Blumenthal2007,Bocquillon1054,Dubois_2013,Kataoka_2013,PhysRevLett.112.226803,Jullien_2014,Ubbelohde2015,PhysRevLett.116.166801}. By applying periodic gate or bias voltages to a nano-scale structure, such as a quantum dot \cite{Kataoka_2013,PhysRevLett.112.226803,Ubbelohde2015,PhysRevLett.116.166801}, a mesocopic capacitor \cite{Feve1169,Bocquillon1054}, or an ohmic contact \cite{Dubois_2013,Jullien_2014}, single electrons can be periodically emitted into a ballistic conductor, leading to an electric current given by the driving frequency times the electron charge \cite{Pekola2013}. \revision{While experiments so far have mainly focused on dynamic sources that emit a single electron per cycle, theoretical works have explored the dynamic generation of more complex quantum states of entangled electrons in normal-metals \cite{PhysRevLett.94.186804,PhysRevB.71.245317,PhysRevB.72.245314,PhysRevB.88.241308}. On the other hand, the concept of controlling the splitting of Cooper pairs with time-dependent driving fields has not been considered before, however, given the very recent experimental progress in the field~\cite{tan2020,PhysRevLett.126.147701,Ranni2021}, such ideas are finally within reach.}
	
In this Letter, we propose and analyze a dynamic Cooper pair splitter that can deliver a noiseless and regular stream of spin-entangled electrons. Specifically, we show how the splitting of Cooper pairs can be controlled by applying time-dependent gate voltages to two quantum dots that are connected to a superconductor. We evaluate the average current and the fluctuations in the output leads and identify the optimal operating conditions for the dynamic Cooper pair splitter to produce a noiseless and regular current. Our proposal seems feasible in the light of recent experimental advances, and it may be realized using current technology.

\begin{figure}[b]
    \centering
    \includegraphics[width=0.97\columnwidth]{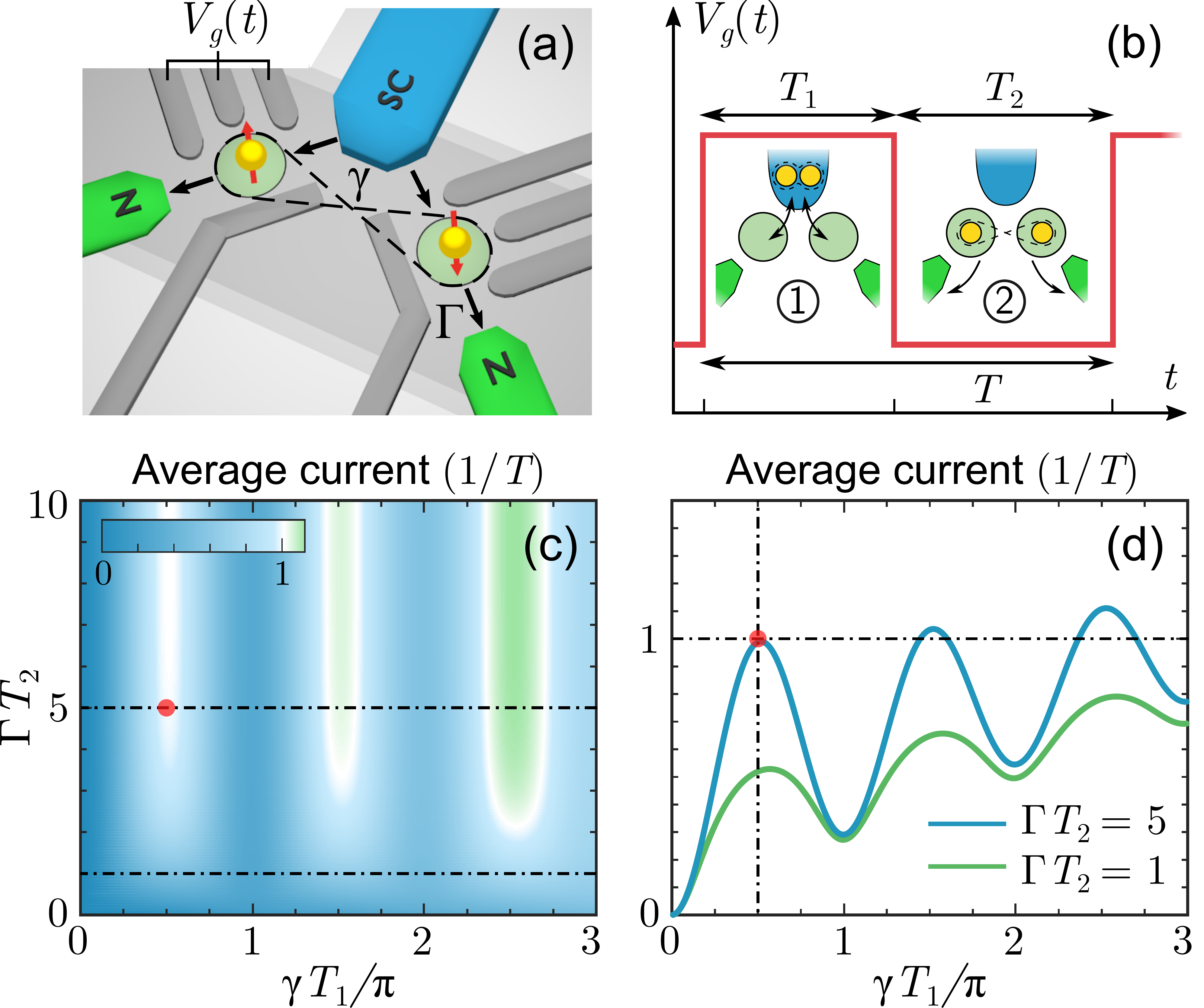}
    \captionsetup{justification=justified,singlelinecheck=false}
    \caption{Dynamic Cooper pair splitter. (a) The Cooper pair splitter consists of two quantum dots (light green) coupled to a superconductor (blue) and two normal-metal drains (green). The splitting of Cooper pairs is controlled with the time-dependent gate voltage, $V_g(t)$. (b) In phase \raisebox{.5pt}{\textcircled{\raisebox{-.9pt} {1}}} of the periodic driving protocol, Cooper pair splitting is tuned into resonance for the time $T_1$, so that a split Cooper pair tunnels into the dots (see insets). In phase \raisebox{.5pt}{\textcircled{\raisebox{-.9pt} {2}}}, Cooper pair splitting is off resonance for the time $T_2$, and the electrons may escape via the drains. (c,d) Average current in the drains as a function of $T_1$ and $T_2$ for $\Gamma = 0.1\gamma$, $\kappa =\gamma$, $\delta =100\kappa$, with $\varepsilon=0$ in  phase \raisebox{.5pt}{\textcircled{\raisebox{-.9pt} {1}}} and $\varepsilon=100\gamma$ in phase \raisebox{.5pt}{\textcircled{\raisebox{-.9pt} {2}}} (see main text for definitions). For the optimal conditions, $\gamma T_1=\pi/2$, and $\Gamma T_2 = 5$, shown with a red dot, one Cooper pair is split per period of the drive.}
    \label{Cooper pair splitter}
\end{figure}

\emph{Dynamic Cooper pair splitter.}--- Figure~\ref{Cooper pair splitter}(a) shows our dynamic Cooper pair splitter consisting of a superconductor coupled to two single-level quantum dots. Cooper pairs from the superconductor are split between the dots due to strong on-site Coulomb interactions, which prevent each dot from being doubly occupied. Electrons on the dots are collected in separate normal-metal drain electrodes. Importantly for our proposal, we dynamically control the splitting of Cooper pairs using a  time-dependent gate voltage $V_g(t)$ as we explain below.

\revision{For energy levels positioned well inside the superconducting gap,} the coherent dynamics of the dots due to the coupling to the superconductor can be described by the effective Hamiltonian \cite{Sauret:Quantum,Eldridge:Superconducting,Hiltscher2011,Walldorf2020,Note1} 
\begin{equation}\label{eq:Heff}
\hat{H}(t)=\!\sum_{\ell\sigma}\epsilon_\ell(t)\hat{d}_{\ell\sigma}^\dagger \hat{d}_{\ell\sigma}^{\phantom\dagger}-\bigg(\gamma\hat d_S^\dagger \!+\!\sum_\sigma\!\kappa \hat{d}_{L\sigma}^\dagger \hat{d}_{R\sigma}^{\phantom\dagger}+\text{H.c.}\!\bigg),
\end{equation}
where $\hat d_{\ell\sigma}^\dagger$ creates an electron with spin $\sigma = \uparrow,\downarrow$ in dot $\ell =L,R$, while $\hat d_S^\dagger \equiv (\hat{d}_{L\downarrow}^\dagger \hat{d}_{R\uparrow}^\dagger-\hat{d}_{L\uparrow}^\dagger \hat{d}_{R\downarrow}^\dagger)/\sqrt{2}$ creates a two-electron spin-singlet state, which is delocalized across the two dots. Here, the (real) amplitudes for Cooper pair splitting and elastic cotunneling are denoted by $\gamma$ and $\kappa$, respectively, and $\epsilon_\ell(t)$ is the time-dependent energy level of each dot \revision{(relative to the chemical potential of the superconductor)}, which we tune by external gates to control the splitting of Cooper pairs and elastic cotunneling between the dots. Elastic cotunneling occurs mainly, when the dot levels are aligned and the detuning $\delta = \epsilon_L-\epsilon_R$ vanishes. Similarly, Cooper pair splitting is on resonance, when the doubly occupied dots have the same energy as the empty dots and the sum $\varepsilon = \epsilon_L+\epsilon_R$ vanishes. In general, Cooper pair splitting and elastic cotunneling lead to coherent oscillations with angular frequencies $\omega_\gamma = \sqrt{4\gamma^2+\varepsilon^2}$ and $\omega_\kappa = \sqrt{4\kappa^2+\delta^2}$, respectively, and both processes are suppressed as $\gamma^2/(\gamma^2+\varepsilon^2/4)$ and $\kappa^2/(\kappa^2+\delta^2/4)$ as we move away from the resonances. These suppression factors provide us with efficient experimental knobs to control the two types of processes. Thus, in the following, we consider the periodic driving protocol in Fig.~\ref{Cooper pair splitter}(b), where Cooper pair splitting is tuned in ($\varepsilon=0$) and out ($\varepsilon\gg\gamma$) of resonance, while elastic cotunneling is kept off resonance ($\delta\gg\kappa$). The duration of each phase is denoted by $T_j$, $j=1,2$, with $\hat H_j$ being the corresponding Hamiltonian, and $T=T_1+T_2$ is the period of the drive.

With large negative voltages on the drains, the time-evolution  is governed by the Lindblad equation \cite{Breuer,PhysRevB.63.165313}
\begin{equation}\label{eq:vonNeumannEq}
\frac{d}{dt}\hat \rho(t)=\mathcal{L}_j\hat \rho(t)=-\frac{i}{\hbar}[\hat H_j,\hat \rho(t)]+\mathcal{D}\hat \rho(t)
\end{equation}
for each of the phases with Liouvillian $\mathcal{L}_j$, $j=1,2$, and  $\hat{\rho}(t)$ is the density matrix of the dots, while the dissipator
\begin{equation}
\mathcal{D}\hat \rho(t)=\!\Gamma\!\!\sum_{\sigma,\ell=L,R}\!\left(\mathcal{J}_{\ell \sigma}\hat \rho(t)-\frac{1}{2}\{\hat \rho(t),\hat d_{\ell\sigma}^\dagger \hat d_{\ell\sigma}^{\phantom\dagger}\}\right)
\label{eq:Dissipator}
\end{equation}
describes the coupling to the drains. The jump operators $\mathcal{J}_{\ell \sigma}\hat \rho(t) \equiv \hat d_{\ell\sigma}^{\phantom\dagger}\hat \rho(t) \hat d_{\ell\sigma}^\dagger$ describe the tunneling of single electrons to the drains with the rate $\Gamma$, and from hereon we take $\hbar, e = 1$. The simple form of the Lindblad dissipator is due to the large negative voltages on the drains. \revisiontwo{Our model can readily be refined according to a specific experiment, but to keep the discussion simple we do not include  additional processes here.}

\emph{Driving protocol.}--- We first consider our driving protocol in the weak coupling limit, $\Gamma \ll \gamma, \kappa$, suppressing elastic cotunneling using the off-resonance condition $\delta \gg \kappa$. In the first phase ($\raisebox{.5pt}{\textcircled{\raisebox{-.9pt} {1}}}$), Cooper pair splitting is tuned into resonance ($\varepsilon = 0$) for the time $T_1$. Ideally, during this phase, a Cooper pair is split between the dots, and we refer to $T_1$ as the loading time. In the second phase ($\raisebox{.5pt}{\textcircled{\raisebox{-.9pt} {2}}}$), Cooper pair splitting is turned off resonance ($\varepsilon \gg \gamma$) for the unloading time $T_2$, allowing the split Cooper pair to leave the dots via the drains. \revision{Since our protocol is based on turning Cooper pair splitting on and off, the step-like scheme is optimal for our purposes, and as shown in Fig.~\ref{Cooper pair splitter}(a), it suffices to tune one of the levels in and out of resonance with the other.} We now evaluate the drain currents to optimize the loading and unloading times.

Figure~\ref{Cooper pair splitter}(c,d) shows the average current for the periodic driving protocol obtained using a method described below. The current oscillates as a function of the loading time $T_1$ due to the coherent oscillations induced by Cooper pair splitting in the first phase. By contrast, the current increases  monotonously as a function of $T_2$ until it saturates for $\Gamma T_2 \gtrsim 5$, reflecting that the unloading time is sufficiently long for the electrons to leave the dots. In this regime, the average current [the blue curve in Fig.~\ref{Cooper pair splitter}(d)] can be captured by the simple expression
\begin{equation}
    I = \frac{1}{T}\left(\sin^2\left(\gamma T_1\right)+ \frac{1}{2}\left[\Gamma T_1+\left(1-e^{-\Gamma T_1}\right) \cos\left(2\gamma T_1\right) \right]\right),
    \label{Average current ideal}
\end{equation}
where the first term stems from the coherent oscillations for very small drain couplings, $\Gamma T_1\ll1$. Corresponding to this term, exactly one Cooper pair is split per period, if $\gamma T_1 = \pi(1/2+m)$ for integer $m$. For finite couplings, the second term describes an unwanted leakage current during the first phase, which can be minimized by choosing a short loading time, $\Gamma T_1\ll 1$. Thus, we find the optimal conditions $\gamma T_1 = \pi/2\gg \Gamma T_1$ and $\Gamma T_2 \simeq 5$ shown with a red dot in Fig.~\ref{Cooper pair splitter}(c,d), where on average one Cooper pair is split per period of the drive.

\begin{figure*}[t]
    \centering
    \includegraphics[width=0.97\textwidth]{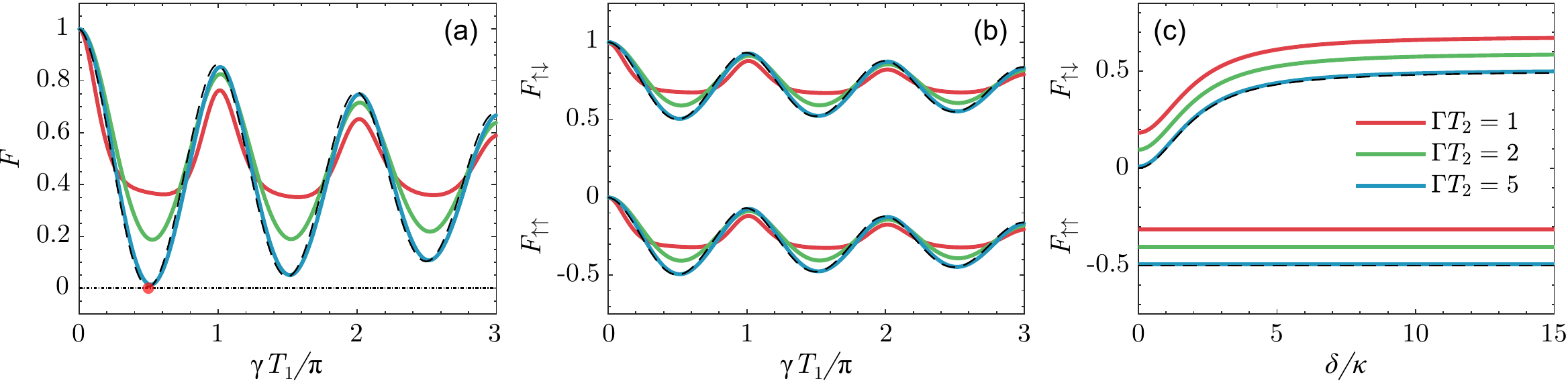}
    \captionsetup{justification=justified,singlelinecheck=false}
    \caption{Noise and spin correlations. (a) Fano factor as a function of the loading time $T_1$ for $\Gamma=0.1\gamma$, $\gamma=\kappa$, $\delta = 100\kappa$, with $\varepsilon=0$ in  phase \raisebox{.5pt}{\textcircled{\raisebox{-.9pt} {1}}}, and $\varepsilon=100\gamma$ in  phase \raisebox{.5pt}{\textcircled{\raisebox{-.9pt} {2}}}. (b) Fano factors of the spin cross-correlations as a function of $T_1$ for the same settings as in panel (a). (c) Fano factors of the spin-correlations as functions of $\delta$ with $\gamma T_1=\pi/2$ and otherwise same parameters as in panel (a). The dashed lines in panels (a), (b), and (c) are given by Eqs.~(\ref{Fano factor}), (\ref{Fano relations}), and (\ref{eq:MGFapp}), respectively.}
    \label{Noise and spin correlations}
\end{figure*}

\emph{Fluctuations.}--- We now proceed with a refined analysis by investigating the noise in the outputs. To this end, we decompose the density matrix as $\hat{\rho}(t) = \sum_{\mathbf{n}} \hat{\rho}(\mathbf{n},t)$, so that $P(\mathbf{n},t) = \text{tr}\left\{ \hat{\rho}(\mathbf{n},t) \right\}$ is the probability that  $\mathbf{n} = (n_L,n_R)$ electrons have been collected in the drains during the time span $[0,t]$ \cite{Plenio1998,Makhlin2001}. The equations of motion for $\hat\rho(\mathbf{n},t)$ are decoupled by introducing counting fields $\boldsymbol{\chi} = (\chi_L,\chi_R)$ via the transformation  $\hat\rho(\boldsymbol{\chi},t) = \sum_{\mathbf{n}}\hat{\rho}(\mathbf{n},t)e^{i\mathbf{n}\cdot \boldsymbol{\chi}}$. We thereby obtain a generalized master equation for $\hat\rho(\boldsymbol{\chi},t)$ with $\chi$-dependent Liouvillians $\mathcal{L}_j(\boldsymbol{\chi})$ by substituting $\mathcal{J}_{\ell \sigma}\rightarrow e^{i\chi_\ell}\mathcal{J}_{\ell \sigma}$ in Eq.~\eqref{eq:Dissipator}~\cite{Walldorf2020}. The moment generating function for the number of emitted electrons after $N$ periods now reads \cite{Pistolesi2004,Albert2010,Potanina2019}
\begin{equation}
    M(\boldsymbol{\chi},N) = \text{tr}\!\left\{\hat\rho(\boldsymbol{\chi},NT) \right\}=\text{tr}\!\left\{ [\mathcal{U}(\boldsymbol{\chi},T,0)]^N \hat\rho_C(0) \right\},
    \label{MGF}
\end{equation}
where the evolution operator for a time-dependent Liouvillian is given by a time-ordered exponential as $\mathcal{U}(\boldsymbol{\chi},t,t_0) = \mathcal{T} \{\exp[\int_{t_0}^t \mathcal{L}(\boldsymbol{\chi},t') dt']\}$, which we can explicitly evaluate for our piecewise constant protocol, and \revision{the cyclic state $\hat \rho_C(t)$} is determined by the eigenproblem
$\mathcal U(\boldsymbol{0},T+t,t) \hat \rho_C(t) = \hat \rho_C(t)$. The zero-frequency current correlators are then given by derivatives of the cumulant generating function $F(\boldsymbol{\chi})=\lim_{N\rightarrow \infty}\ln[M(\boldsymbol{\chi},N)]/NT$  as $\langle \!\langle I_L^k I_R^l\rangle\!\rangle = \partial_{i \chi_L}^k\partial_{i \chi_R}^l F(\boldsymbol{\chi})|_{\boldsymbol{\chi}=0}$. With these definitions, we can calculate the currents and their correlations using the methods from Refs.~\cite{refId0,PhysRevLett.100.150601,PhysRevB.82.155407} and obtain simple expressions as those in Eq.~\eqref{Average current ideal} and Eq.~\eqref{Fano factor} below.

\emph{Noise and Fano factor.}--- Figure~\ref{Noise and spin correlations}(a) shows the Fano factor, $\revision{F} =\langle\!\langle I_\ell^2\rangle\!\rangle/\langle\!\langle I_\ell\rangle\!\rangle$, $\ell=L,R$, of the noise in the drain electrodes, which for regular transport is suppressed below the Poisson value of one. Here, the Fano factor oscillates as a function of the loading time $T_1$, similarly to the average current in Fig.~\ref{Cooper pair splitter}(c,d), and for long unloading times, $\Gamma T_2\gg 1$, we find the simple expression
\begin{equation}
F = 1-TI+\frac{2\Gamma T_1(1+\Gamma T_1)+e^{-2\Gamma T_1}-1}{8 TI},
 \label{Fano factor}
\end{equation}
corresponding to the dashed line in Fig.~\ref{Noise and spin correlations}(a). For weak drain couplings, $\Gamma T_1\ll1$, the Fano factor reduces to $F=\cos^2(\gamma T_1)$, reflecting the coherent oscillations in the first phase. For larger couplings, the leakage current in the loading phase generates noise since more than one Cooper pair may be split during each period. Thus, we may minimize the noise by choosing $\gamma T_1 = \pi/2$ together with a long unloading time, $\Gamma T_2 \gg 1$, corresponding to the red dot in Fig.~\ref{Noise and spin correlations}(a).  There, the device produces exactly $N$ split Cooper pairs after $N$ periods, and it is noiseless in contrast to a static Cooper pair splitter. \revisiontwo{Deviations from the optimal conditions increase the noise due to cycle-missing events, in which no Cooper pair is split, however, the entanglement of each split pair is unaffected.}

\emph{Spin-current correlations.}--- Next, we consider the cross-correlations between the spin currents in each drain, which play an important role for detecting the entanglement of the split Cooper pairs \cite{Kawabata2001,Malkoc_2014,PhysRevB.96.064520,Brange2017}. It it straightforward to include spin-dependent counting fields in Eq.~(\ref{MGF}), and in Fig.~\ref{Noise and spin correlations}(b) we show the resulting cross-correlations, $F_{\sigma\sigma'} = \langle\!\langle I_{L\sigma} I_{R\sigma'}\rangle\!\rangle/\sqrt{\langle\!\langle I_{L\sigma}\rangle\!\rangle\langle\!\langle I_{R\sigma'}\rangle\!\rangle}$, as functions of the loading time $T_1$. The anti-parallel spin currents are positively correlated, while parallel spins exhibit negative correlations, as expected for a split Cooper pair in a spin-singlet state. For long unloading times, $\Gamma T_2 \gg 1$, we find that the spin correlations Fig.~\ref{Noise and spin correlations}(b) can be related to the Fano factor of the charge currents as
\begin{equation}
    F_{\uparrow \downarrow} =1+F_{\uparrow \uparrow} = \frac{1}{2}\left(F+1\right),
    \label{Fano relations}
\end{equation}
making it possible to determine the spin correlations from noise measurements of the charge currents in the drains.

\begin{figure*}[t]
    \centering
    \includegraphics[width=0.97\textwidth]{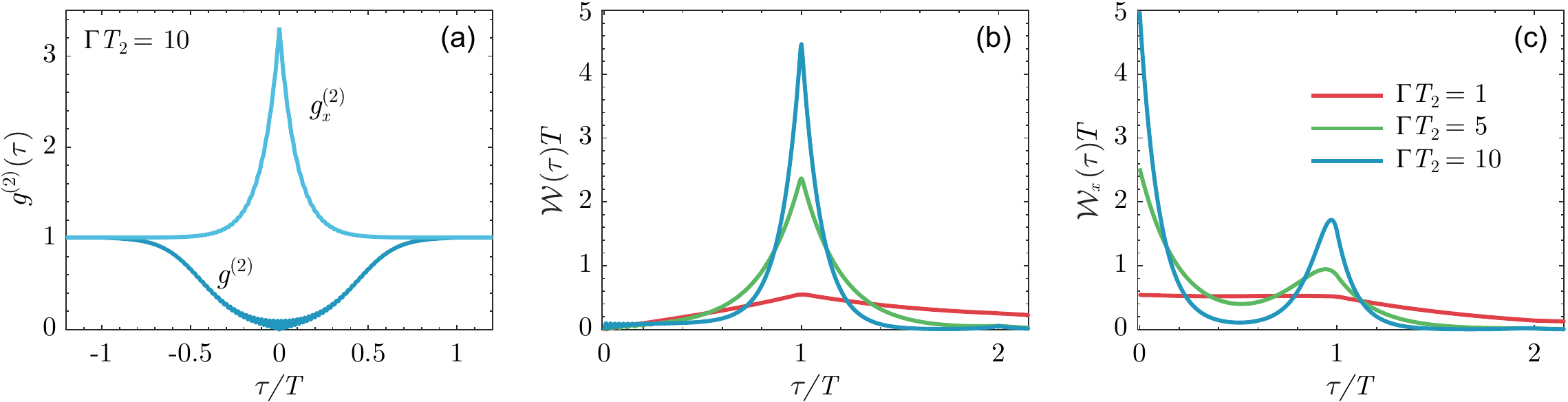}
    \captionsetup{justification=justified,singlelinecheck=false}
    \caption{Time-domain observables. (a)  Auto and cross-correlations ($g^{(2)}_x$) of the drain currents. (b) Distributions of waiting times between electrons tunneling into the same drain. (c) Distributions of waiting times between electrons tunneling into different drains. Parameters are $\gamma T_1 = \pi/2$, $\Gamma = 0.05\gamma $, $\kappa = \gamma$,  $\delta = 20\kappa$ with  $\varepsilon=0$ in phase \raisebox{.5pt}{\textcircled{\raisebox{-.9pt} {1}}} and $\varepsilon=20\gamma$ in phase \raisebox{.5pt}{\textcircled{\raisebox{-.9pt} {2}}}.}
    \label{g2 and waiting time distributions}
\end{figure*}

In Fig.~\ref{Noise and spin correlations}(c), we consider the spin-correlations as we tune elastic cotunneling into resonance. The negative correlations for parallel spins are essentially unchanged as elastic cotunneling is included. On the other hand, the positive correlations for anti-parallel spins are gradually washed out by elastic cotunneling. We also consider a short loading time, $\Gamma T_1 \ll 1$, so that the leakage current is negligible during the first phase, and a long unloading time, $\Gamma T_2 \gg 1$, ensuring that at most one Cooper pair is split per period. Each uncorrelated period may then produce one of five outcomes; either no Cooper pair is split and no electrons reach the drain, or one Cooper pair is split and the two electrons tunnel into either of the two drains. We then approximate the generating function as
\begin{equation}
\begin{split}
M(\boldsymbol{\chi},N) \simeq  \bigg[&1-p\!+\!\frac{pq}{2} \!\left(e^{i(\chi^L_\uparrow+\chi^L_\downarrow)}\!+\!e^{i(\chi^R_\uparrow+\chi^R_\downarrow)}\right)\\
+&\frac{p(1-q)}{2}\!\left(e^{i(\chi^L_\uparrow+\chi^R_\downarrow)}\!+\!e^{i(\chi^R_\uparrow+\chi^L_\downarrow)}\right)\!\bigg]^N,
\end{split}
\label{eq:MGFapp}
\end{equation}
where $p = \sin^2(\omega_\gamma T_1/2)\gamma^2/(\gamma^2+\varepsilon^2/4)$ is the probability that a Cooper pair is split in the first phase, and $q = \frac{1}{2}\kappa^2/[\kappa^2+(\Gamma^2+\delta^2)/4]$ is the probability that the electrons leave via the same drain in the second phase due to elastic cotunneling. In Fig.~\ref{Noise and spin correlations}(c), we show calculations of the spin-correlations based on this approximation and find good agreement with the exact results. Furthermore, we find $F_{\uparrow \uparrow} = -p/2$ and $F_{\uparrow \downarrow} = 1-p/2-q$, which shows that, in this regime, $F_{\uparrow \uparrow}$ and $F_{\uparrow \downarrow}$ are sufficient to determine $p$ and $q$ and thus fully characterize the noise statistics.

\emph{Time-domain analysis.}--- Above, we focused on conventional low-frequency measurements.  However, as in the experiment of Ref.~\cite{Ranni2021}, more information can be obtained by analyzing the fluctuations in the time-domain. We thus consider the $g^{(2)}$-function of the  currents~\cite{PhysRevA.39.1200,PhysRevB.85.165417}
\begin{equation}
    g^{(2)}_{\ell \ell'}(\tau) =\int_0^T dt_0  \frac{\langle\!\langle \mathcal{J}_{\ell} \mathcal{U}(\mathbf{0},t_0+\tau,t_0)\mathcal{J}_{\ell'} \rangle\!\rangle_{t_0}}{\langle\!\langle \mathcal{J}_{\ell} \rangle\!\rangle_{t_0+\tau}\langle\!\langle \mathcal{J}_{\ell'} \rangle\!\rangle_{t_0}}P_{\ell'}(t_0)
\end{equation}
\revision{with $\mathcal{J}_\ell \equiv \sum_{\sigma}\mathcal{J}_{\ell \sigma}$}. Here, the probability density for the time that a tunneling event occurs reads $P_\ell(t) = \langle\!\langle \mathcal{J}_{\ell}\rangle\!\rangle_t / \int_0^T d\tau \langle\!\langle \mathcal{J}_{\ell}\rangle\!\rangle_\tau$, and $\langle\!\langle \mathcal{A}\rangle\!\rangle_t = \text{tr}\{ \mathcal{A} \hat \rho_C(t)\}$ is the expectation value of $\mathcal{A}$. In this definition, the correlations due to the periodic drive have been factored out.
 
Figure~\ref{g2 and waiting time distributions}(a) shows the  the auto ($g^{(2)}$) and  cross-correlations ($g_x^{(2)}$) of the drain currents. The auto-correlations are suppressed around $\tau =0$, corresponding to the anti-bunching of electrons due to the Coulomb interactions on the dots. The cross-correlations, by contrast, exhibit a peak well above the uncorrelated value of one, showing how Cooper pair splitting leads to nearly simultaneous tunneling of electrons into different leads.
 
Information about the regularity of the dynamic Cooper pair splitter can be obtained from the waiting times between the tunneling events~\cite{Brangeeabe0793,Ranni2021}. The distribution of waiting times can be expressed as \cite{Brandes:Waiting,Albert2011,potanina2017,Walldorf2018,Wrzesniewski2020}
\begin{equation}
   \mathcal{W}_{\ell\ell'}(\tau) =\int_0^T dt_0\frac{\langle\!\langle \mathcal{J}_{\ell} \mathcal{U}_\ell(t_0+\tau,t_0)\mathcal{J}_{\ell'} \rangle\!\rangle_{t_0}}{\langle\!\langle \mathcal{J}_{\ell'} \rangle\!\rangle_{t_0}}P_{\ell'}(t_0),
\end{equation}
where $\mathcal{U}_\ell(t,t_0) = \mathcal{T}\exp[\int_{t_0}^t \left(\mathcal{L}(\mathbf{0},t')-\revision{\mathcal{J}_\ell}\right) dt']$ is the time-evolution operator excluding electron tunneling into drain $\ell$. For $\ell = \ell'$, we obtain the distribution of waiting times between electrons tunneling into the same lead, while for $\ell \neq \ell'$, the distribution characterizes the waiting time between electrons tunneling into different leads.

Figure~\ref{g2 and waiting time distributions}(b,c) shows both types of distributions. With a long unloading time, $\Gamma T_2 \gg 1$, electrons tunneling into the same drain tend to be separated by the period of the drive due to the regular splitting of Cooper pairs. For shorter unloading times, the dots are not always emptied in the second phase, and the Cooper pair splitting becomes less regular. This picture is corroborated by the distributions for tunneling into different leads in Fig.~\ref{g2 and waiting time distributions}(c). Here, the first peak at short waiting times corresponds to the tunneling of electrons from the same split Cooper pair into different leads. In addition, the second peak corresponds to the waiting time between the last electron from one split Cooper pair and the first electron from the next split pair, and those are spaced by an interval, which is slightly shorter than the driving period.

\revision{\emph{Experimental perspectives.}---}  \revision{To provide realistic parameters for our setup, we note that the superconducting gap can be around $\Delta \simeq 200$~$\mu$eV \cite{tan2020} (corresponding to a temperature of 2.3~K), while the amplitudes for Cooper pair splitting and elastic cotunneling can be around $\gamma,\kappa\simeq 10$~$\mu$eV. With a temperature of 100 mK, quasiparticle tunneling from outside the gap is then strongly suppressed. Moreover, with a detuning of $\delta = 100$~$\mu$eV, elastic cotunneling can be suppresed by a factor of $\kappa^2/(\kappa^2+\delta^2) = 0.01$ compared to Cooper pair splitting. In addition, with tunneling rates being $\Gamma \simeq 1$ $\mu$eV, we find $T_1 \simeq 1$~ns for the loading phase and $T_2 \simeq 10$~ns for the unloading phase, and the optimal driving frequency is then about $100$~MHz with currents of about 10 pA.}

\revision{\emph{Conclusions.}}--- We have proposed a dynamic Cooper pair splitter that can generate a noiseless and regular flow of spin-entangled electrons, when operated under optimal conditions. Our proposals appears feasible in the light of recent experiments \cite{Brangeeabe0793,Ranni2021}, and it may thus pave the way for the controlled generation of spin-entangled electrons. \revisiontwo{Moreover, additional control of the electron spins may be achieved with more elaborate pulse sequences~\cite{Petta2005}, for instance by using materials with strong spin-orbit coupling to generate effective, time-dependent magnetic fields~\cite{Flindt2006}. Finally,} entanglement witnesses based on current cross-correlations \cite{PhysRevB.96.064520,Brange2017} may be used to certify the spin-entanglement of the split Cooper pairs.

\acknowledgements
\emph{Acknowledgements.}--- We thank N.~Walldorf for his involvement at an early stage of the project and A.~Ranni and V.~F.~Maisi for useful discussions. We acknowledge support from Aalto Science Institute and Academy of Finland through the Finnish Centre of Excellence in Quantum Technology (project numbers 312057 and 312299) and grants number 308515 and 331737.


%

\end{document}